\documentclass[11pt,a4paper]{article}
\usepackage{jcappub,aas_macros}

\title{Microlensing due to both gravitation and refraction as a further
probe of universe evolution}
\author[a]{G. Cavalleri}
\author[b]{F. Barbero}
\author[a]{E. Tonni}
\author[c]{and S. Covino}
\affiliation[a]{Universit\`a Cattolica, via Musei 41, 25121 Brescia, Italy}
\affiliation[b]{Institute for Foundational Studies Hermann Minkowski, Montreal, Canada}
\affiliation[c]{INAF/Brera Astronomical Observatory, via Bianchim 46, 23807, Merate (LC), Italy}

\emailAdd{giancarlo.cavalleri@gmail.com}

\date{}

\abstract{Microlensings events are predicted for the light coming from cosmological sources. In addition to the microlensing due to gravitation lensing, microlensing produced also by refraction of light due to either ionized, or not, gas clouds can be considered. A detailed prediction is here given assuming that the ray
of light coming from the distant source traverses a gas cloud with a King's
density profile for various possible environments. We conclude that the additional deviation due to relativistic refraction is in most cases negligible compared to the gravitational deviation. Deviation due to refraction can anyway become an interesting analysis tool for future facility with great resolving power and the effects can be singled out with dedicated surveys.}


\begin{document}
\maketitle

\section{Introduction}
 \label{s:1}

Microlensing is a topical subject, and regards variations of intensity of the received light from distant sources due to objects along the line of sight \citep[e.g.][]{1998ApJ...509L..41D}.  It is not typically observable as a splitting of an image (or of an image and one or more arcs), because deviations are typically between $0.0005^{\prime \prime }$ and $0.01^{\prime \prime }$ \citep{1998LRR.....1...12W}. In fact, such microlensing is due to deviations of light not only caused by gravitation, but also by refraction of light due to either ionized, or not, gas clouds. We here compute the possible importance of microlensing due to refraction in various situations and evaluate its importance compared to gravitational microlensing.

The paper is organized as follows: in Sec. \ref{s:2}, we summarize the expressions of the refraction index $n_{r}$ for both ionized and neutral hydrogen. Since $n_{r}$ depends on the concentration $N$ of hydrogen, we consider the typical density profiles of galaxies, galaxy clusters, HII regions, giant molecular clouds, dense clouds, protostars (PSs), and regular stars in the main sequence (MSSs). Einstein Theory of Gravity (ETG) is known to be certainly valid up to, and including, the second order, the so called post-post-Newtonian (PPN) approximation. However, even limiting to the first order approximation, called linearized Einstein theory or post-Newtonian (PN) approximation, the computations to derive the deviation due to microlensing when the lensed source is at cosmological distance, i.e. its recession velocity is an appreciable fraction of the speed of light, is still difficult. In general it requires the calculation of the whole trajectory and that is why, in the literature, at best of our knowledge, we have not found the calculations of gravitational deviations of light beams traversing a given density profiles. In Sec. \ref{s:3}, we show that the required computations can effectively and more easily be
performed in the framework of an alternative gravity theory sketched in the Introduction (\ref{s:1}), at the end of Sec. \ref{s:3}, in the Conclusion (\ref{s:5}), and in Appendix A of \citet{2013PhRvE..87d3202C}. Such alternative theory has been shown in \citep{2013PhRvE..87d3202C} to lead to same same predictions of Einstein theory at first order approximation, sufficient to calculate the deviation of light due of gravitation. We are then able to provide a remarkably compact final expression, valid for both the gravitational deviation, and, with the only change of a proportionality constant, for the deviation due to refraction in media \citep{2013PhRvE..87d3202C}. In Sec. \ref{s:4} we obtain the explicit expression of the static deviation $\Delta _{0}$ (i.e., corresponding to negligible redshift $z$) due to both gravitation and refraction, relevant to a gas cloud with spherical symmetry). We conclude in Sec. \ref{s:5} commenting our results.

\section{Neutral and ionized hydrogen refraction indexes $n_{r}$, and possible candidates to obtain appreciable deviations}
\label{s:2}

In order to calculate deviations of electromagnetic rays because of refraction, we must know the speed of light $|\mathbf{c}_{0}(\mathbf{r})|=c\,n_{r}(\mathbf{r})$,
where $n_{r}$ is the refractive index of the hydrogen surrounding the receding object, function of the position $\mathbf{r}$. To find $n_{r}(\mathbf{r})$ we separately consider both ionized and neutral hydrogen.

The existence of a filamentary, low-density intergalactic medium (IGM), which contains the bulk of the hydrogen in the universe, is predicted to be
a product of primordial nucleosynthesis \citep{1995Sci...267..192C}. Indeed, the application of the Gunn-Peterson effect \citep{1965ApJ...142.1633G} allows one to infer
that the hydrogen component of the diffuse IGM is highly ionized by $z\sim5$ \citep{1991AJ....101.2004S}. Moreover, from QSO absorption studies we also know that
neutral hydrogen accounts for only a small fraction ($\sim10\%$) of the nucleosynthetic baryons at early epochs \citep{1995ApJ...440..435L}.

A large amount of relatively dense ionized gas is present in galaxy clusters. The depths of their potential wells keep the gas temperature around $10^{7}$ K and the gas is almost totally ionized. Of course there are also recombination lines produced by metals, as iron, etc. Their percentage in mass with respect to hydrogen is $\sim 1\%$ \citep{1997ApJ...488...35R}, about half the solar percentage. The amount of ionized gas is relevant, being of the order of 10-20\% the total mass locked in stars and therefore forming the galaxies in a galaxy cluster \citep{1998ApJ...503..518F}. The number density\textsl{\ }$N$\textsl{\ }per unit volume is however low, of the order \citep{1997ApJ...488...35R} $N\approx 10^{-4}$ particles/cm$^{3}$. In some case it is possible to reach the value $N\approx 10^{-3}$ cm$^{-3}$. To be conservative, we use the latter value that is the maximum one. The corresponding plasma angular frequency is 
\begin{equation}
\omega _{p}=e\left( \frac{4\pi N}{m_{e}}\right) ^{1/2}\approx 10^{3}s^{-1}\ ,
\text{ }  \label{eq:1}
\end{equation}
where $e$ and $m_{e}$ are the electron charge and mass, respectively.

The corresponding refractive index is: 
\begin{equation}
n_{r}(\mathrm{cluster})=[1-(\omega_{p}/\omega)^{2}]^{1/2}\ .  \label{eq:2}
\end{equation}
Yellow light has $\omega=\omega_{y}\simeq3\times10^{15}$ s$^{-1}$ so that: 
\begin{equation}
n_{ry}(\mathrm{cluster})\simeq1-\frac{1}{2}\left( \frac{\omega_{p}}{
\omega_{y}}\right) ^{2}\approx1-10^{-25}\ .  \label{eq:3}
\end{equation}
The difference from unity is so small that no appreciable deviation of light can be detected.

Inside spiral galaxies, ionized hydrogen is present in the HII regions, which are only 2\% of the projected area of a galaxy and are close to the
galactic plane of symmetry \citep{1989agna.book.....O}. The refraction index is less than unity and differs very little from $1$. In fact, the maximum gas density in
the HII regions is $N\approx 1$ cm$^{-3}$, so that $\omega _{p}\simeq5.7\times 10^{4}$ s$^{-1}$. Consequently: 
\begin{equation}
n_{ry}(\mathrm{region\ HII})\simeq 1-\frac{1}{2}\left( \frac{\omega _{p}}{
\omega _{y}}\right) ^{2}\approx 1-10^{-22}\ ,  \label{eq:4}
\end{equation}
thus corresponding to a non-detectable deviation. We can therefore conclude that there are no appreciable deviations of ray beams due to ionized gas in
the universe.

Let us now consider neutral hydrogen. The gas density in a star, a galaxy, or a galaxy cluster can vary within a wide range: typical values for the
density in the galactic disk is $0.1\div 100$ cm$^{-3}$, while in some denser regions (i.e., star forming regions, molecular clouds, etc.) we can
have $10^{4}\div 10^{8}$ cm$^{-3}$. In a solar--type star atmosphere the number density $N$ is of the order of $10^{17}$ cm$^{-3}$ close to the
visible layers \citep{1990hsaa.book.....Z} while in the extragalactic medium $N$ is lower than $10^{-12}$ cm$^{-3}$ \citep{1988asco.book.....H}.

The refractive index for the spectral line D of sodium in neutral hydrogen at atmospheric pressure ($\sim 10^{5}$\thinspace Pa) and at a temperature $20
{{}^{\circ }}$C ($\sim 293$K) is\footnote{{\tiny http://hypertextbook.com/physics/waves/refraction/index.shtml}} $\left.n_{r(lab)}\right\vert _{D}=1+1.32\times 10^{-4}$. In the infrared, with $\lambda =8\mu m$, it is $\left. n_{r(lab)}\right\vert _{8\mu m}=1+1.38\times10^{-4}$ with a difference of \ $\sim 3\%$\ with respect to the value for
the D line of sodium. Being this fractional difference smaller than the uncertainty in the number density $N$, we can take a single intermediate
value for both visible light and near infrared: 
\begin{equation}
n_{r(lab)}=1+1.33\times 10^{-4}\ ,  \label{eq:49}
\end{equation}
relevant to a hydrogen numerical density $N_{lab}\simeq 2.5\times 10^{19}$ cm$^{-3}$. Since $n_{r}-1$ scales linearly with the numerical density $N$, we
obtain the general relation: 
\begin{equation}
\frac{n_{r}-1}{n_{r(lab)}-1}=\frac{N}{N_{lab}}\ .  \label{eq:50}
\end{equation}%
We derive from equations (\ref{eq:49}) and (\ref{eq:50})
\begin{equation}
n_{r}=1+5.28\times 10^{-30}N\text{ .}  \label{eq6bis}
\end{equation}

\subsection{Matter density profiles}

The correct description of the matter density profiles for a galaxy, a galaxy cluster, and for the external layers of the stars is a quite
complicated issue. Within the context of the present paper we can however refer to a simplified and analytic approach still able to correctly give us
the right order of magnitude of the quantities we are estimating.

The galaxy or galaxy cluster density profile is often adequately described by the modified King approximation to the isothermal sphere \citep{1988xrec.book.....S}: 
\begin{equation}
N(r)=N(0)\left[ 1+\left( r\left/ r_{C}\right. \right) ^{2}\right] ^{-3b/2}[%
\text{cm}^{-3}]\ ,  \label{eq:47}
\end{equation}%
$N(0)$ being the central concentration, $r_{C}$ the core radius, $r$ the distance from the center of the galaxy or galaxy cluster, $b=\mu m_{p}\sigma
_{r}^{2}/(kT)$, where $\mu $ is the mean molecular mass (in atomic units), $m_{p}$ the proton mass, $k$ the Boltzmann constant, $T$ the gas temperature,
and $\sigma _{r}^{2}$ the line-of-sight velocity dispersion. Typical values for the best observed galaxy clusters \citep{1984ApJ...276...38J,1999ApJ...511...65J} are $b=2/3$, $r_{c}=250$\thinspace kpc, and $N(0)=10^{-2}$m$^{-3}$, corresponding to a mass density $3\times 10^{14}$\thinspace\ M$_{\odot }$/M\thinspace pc$^{3}$ \citep{1999ApJ...511...65J}.

Larger concentrations $N$ of gas molecules are those of galaxies and of galaxy cores. The average values in the galactic plane of symmetry are $N\simeq 1$ cm$^{-3}$, still highly insufficient to give measurable deviations through refraction.

The deviations due to refraction are insufficient also for the giant molecular clouds, with radii $\sim 20$ pc, masses $\sim 4\times10^{5}M_{\odot }$, hence average $N\simeq 10^{3}$ cm$^{-3}$. The total mass of the giant molecular clouds in a spiral galaxy is $\simeq 2\times 10^{9}M_{\odot }$, so that their number in a galaxy is $5\times 10^{4}$ \citep[][Table\,1]{2004A&A...415..171V}.

Other candidates are the dense clouds, with radii $\sim 1$ parsec, masses $\sim 10^{3}M_{\odot }$, hence average $N\sim 10^{4}$cm$^{-3}$. Their
total mass in a spiral galaxy is $\sim 5\times 10^{8}M_{\odot }$, so that their number in a galaxy is $\sim 10^{5}$ \citep{2002MNRAS.330..583L,2004A&A...415..171V}. In the cores of
the dense clouds we may have $N\sim 10^{5}$cm$^{-3}$, which is still too small to give a measurable deviation due
to refraction.

The best candidates to have measurable deviations are actually PSs in the stage just following the \textquotedblleft T Tauri star\textquotedblright\ until
becoming a pre-main-sequence star \citep{2007ARA&A..45..565M}. The density profile of a PS in the stage of evolution can still be taken of the King type with $b=2.7$,
so that the exponent of equation \ref{eq:47} becomes $1.8$, and the relation between the PS mass $M$ and its core radius $r_{C}$ turns out to be $M=17.9N(0)m_{p}r_{C}^{3}$, 
where $m_{p}$ is the proton mass. If we take $M=M_{\odot }$ \ and $N(0)=10^{17}$cm$^{-3}$, we obtain 
\begin{equation}
r_{C}\approx 8.75\times 10^{12}\text{ [cm]}.  \label{9}
\end{equation}
The density profile is therefore: 
\begin{equation}
N(r)=10^{17}\left[ 1+\left( r\left/ r_{C}\right. \right) ^{2}\right] ^{-1.8}[
\text{cm}^{-3}]\ ,  \label{9b}
\end{equation}
where $r$ is the distance from the center of the PS. The relevant refraction coefficient consequent to equations (\ref{eq6bis}) and (\ref{9b}) is: 
\begin{equation}
n_{r}=1+5.28\times 10^{-7}\left[ 1+\left( r\left/ r_{C}\right. \right) ^{2}
\right] ^{-1.8}\text{, }  \label{eq:10}
\end{equation}
with $r_{C}\approx 8.75\times 10^{12}$cm.

The deviation due to a PS, and the probability that a relevant microlensing be observable, is calculated in Sec. \ref{s:4a}. In Sec. \ref{s:4b} we
calculate the deviation due to the atmosphere of a MSS, and also the probability to observe such an event. For simplicity, the atmosphere of a
MSS can be modeled as an exponential decrease from the MSS photosphere. We can write \citep{1984avis.book.....B}: 
\begin{equation}
N(r)=N\left( r_{p}\right) \exp \left( \frac{r_{p}-r}{r_{H}}\right) \ ,
\mbox{
for }r>r_{p}\ ,  \label{eq:48}
\end{equation}
$r_{p}$ being the radius of the photosphere, and $r_{H}=kT/\mu m_{p}g$ the density scale height, with $\mu $ as the atomic mass in units of the proton
mass $m_{p}$, and $g$ as the surface gravity. The atmosphere of a MSS is ionized, so that the refraction index is expressed by equations (\ref{eq:1})
and (\ref{eq:2}).

\section{Gravitational deflection when light traverses a King's density profile \label{s:3}}

The calculation of the gravitational lens effect when the light traverses a gas cloud with a King's density profile
is very difficult if one uses the General Relativity (GR), because the whole trajectory has to be obtained.
That is probably why it is not reported, at best of our knowledge, in literature. Here we use instead the alternative gravitational theory found in 
\citet{2013PhRvE..87d3202C} that allows one to obtain the deviation $\Delta _{z}$ without knowing the whole trajectory of the light
ray. The alternative theory leads to the same first order expansion of Einstein's theory, called post Newtonian approximation (PN), which is sufficient to
calculate the deviation of EM rays due to gravitation. The speed of light is considered as variable, with a refractive index given, at first order, by Eq.
(72) of \citet{2013PhRvE..87d3202C}. That property implies a second advantage when there is medium (in our case a gas), because it is simple to obtain an equivalent compound refractive index, thus allowing the calculation of the deviation due to both gravitation and usual refraction. 

A particular comment is however proposed here regarding the relativistic calculations performed by \citet{2013PhRvE..87d3202C}, and used in the present paper. In fact, those calculations have been carried out by special relativity (SR) instead of GR. Cosmological observations, in particular the spatial fluctuations of the cosmic background radiation, lead to the flat, pseudo-Euclidean space-time of SR for the average \textquotedblleft dust" universe, curvature being localized around massive objects \citep{2000Natur.404..955D}. In any case, gravitation can be treated in the flat space-time of SR starting from a Lagrangian with a second rank tensor for the potential \citep{1975PhRvD..12.2200C}, and then applying an iterative procedure (because the theory is infinitely nonlinear) that converges to GR even with
the most general gauge, as shown by \citet{1975PhRvD..12.2203C}. The fact that the potential has to be a second-rank tensor has been subsequently proved by \citet{1983NCimB..75...50C}. Gravitation acts even on meter sticks and clocks, so that the space-time measured by these \textquotedblleft renormalized" units turns out to be Riemannian. It is also possible to use semi-renormalized units \citep{1977PhRvD..15.3065C}, or even to remain in flat-space and retain increasing orders of the resultant series (in $\varphi \left/ c\right. $, where $\varphi $ is the scalar gravitational potential and $c$ the speed of light) due to the iterative procedure.

Another seeming problem regards the choice of the reference frame. Cosmologists are so accustomed to comoving coordinates (leading to the Robertson-Walker metrics), that they always consider the receding galaxies as at rest, and only speak about proper, or peculiar, motion with respect to the local, comoving coordinates. However, it was just GR that assured one that the choice of the coordinate system is arbitrary. One can therefore take a frame $S$ at rest with the Earth $E$,
and it is in this reference $S$ that the gravitational lens effect has been calculated when the deflector is in motion \citep{2005MNRAS.359L..19S}. The use of the frame $S$ (with noncomoving coordinates) is convenient to calculate a possible contribution to the gravitational lens effect due to refraction. In this case there is a third reason that allows one to use SR (besides the zero curvature of our present universe, and the possibility to treat gravitation by means of SR). In fact, the deviation due to refraction is a small fraction (as turns out) of the gravitational lens effect, so that one can use the small perturbation theory. The cosmological deviations in the case the space was not flat are already taken into account in the
gravitational lens effect. We can therefore use SR to deal with the refraction of light as done in \citet{2013PhRvE..87d3202C}. The deviation of an electromagnetic ray due to the refraction when a density enhancement is moving with respect to the observer indeed has never been considered, perhaps because until 1996 there was no treatment when the fluids and the interface are moving, even at nonrelativistic speeds \citep[see][]{1996PhRvE..54.6291A}. The relativistic version of \citet{1996PhRvE..54.6291A} has been performed by \citet{2013PhRvE..87d3202C}.

We use here the following expression for the refractive index, whose first order expansion coincides with Eq. (72) of \citet{2013PhRvE..87d3202C}, but is
more accurate at the second order: 
\begin{equation}
n_{G}=\left[ 1+4\chi \left( r\right) \left/ c^{2}\right. \right] ^{1\left/
2\right. }\text{.}  \label{p14}
\end{equation}
The scalar gravitational potential $\chi \left( \mathbf{r}\right) $, function of the position $\mathbf{r}$, is given by the Newtonian expression: 
\begin{equation}
\chi \left( r\right) =\int_{\infty }^{r}\mathbf{E}_{g}\left( \mathbf{r}
^{\prime }\right) \cdot d\mathbf{r}^{\prime }\text{,}  \label{p15}
\end{equation}
$\mathbf{E}_{g}$ being the Newtonian gravitational field: 
\begin{equation}
\mathbf{E}_{g}\left( \mathbf{r}^{\prime }\right) =-\frac{\mathbf{\hat{r}}
^{\prime }}{r^{\prime \text{ }2}}G\int_{0}^{r^{\prime }}\rho \left( r\right)
4\pi r^{2}dr\text{ },  \label{p16}
\end{equation}
where $\rho $ is the mass density. Once we have the expression of the refractive index $n_{G}$, given by equation (\ref{p14}) as a function of the
distance $r$ from the center of a spherical distribution, equation (\ref{A7}) of Appendix \ref{s:A} allows one to calculate the static (i.e., for a
deviating gas cloud at rest with the observer) deviation angle $\Delta _{0}$ without knowing the trajectory.

Once obtained the static deviation $\Delta _{0}$ by means of equation (\ref{A7}) of Appendix \ref{s:A}, the corresponding relativistic expression is
given, in exact form, by equation (65) of \citet{2013PhRvE..87d3202C}. For our purposes, it is sufficient to use its first order approximation in $\beta =v/c$, given by equation (63) of \citet{2013PhRvE..87d3202C} that we report here, at the same time adding its expression as a function of $z$: 
\begin{equation}
\Delta _{\beta }=\Delta _{0}\sqrt{\frac{1+\beta }{1-\beta }}=\Delta
_{0}(z+1)=\Delta _{z}\text{ .}  \label{50}
\end{equation}%
The above relativistic dependence, originally found in \citet{2013PhRvE..87d3202C} for pure refraction of light in gases, is formally equal to the one due to gravitation calculated by \citet{2004PhRvD..69f3001W} in the case of a single concentrated
mass using the linearized Einstein theory. The difficulty of the latter one concerns the calculation of $\Delta _{0}$\ for a diffused mass, as is the
case of the King's profile.

Notice that Eq. (\ref{50}) gives the total deviation of the EM ray
\begin{equation}
\Delta _{z}=\Delta _{\beta }=\vartheta _{Q}+\vartheta _{E}\text{ ,}
\label{51}
\end{equation}
where $\vartheta _{Q}$\ and $\vartheta _{E}$\ denote the angles between the
non deviated ray and the deviated ray near the lensed source and the Earth,
respectively. The deviating gas cloud has a distance\ from the Earth proportional
to $\beta =v/c$, while the one of the lensed source at cosmological distance is proportional to $\beta _{Q}$
we take close to $1$. Consequently, 
\begin{equation}
\frac{\vartheta _{E}}{\vartheta _{Q}}=\frac{1-\beta }{\beta }\text{ .}
\label{52}
\end{equation}
We derive from Eqs. (\ref{50})-(\ref{52})
\begin{equation}
\vartheta _{E}=\Delta _{\beta }(1-\beta )=\Delta _{0}\sqrt{1-\beta ^{2}}
=\Delta _{0}\frac{\sqrt{1+z+z^{2}}}{1+z+z^{2}/2}\text{ .}  \label{53}
\end{equation}

\section{Light's deviations due to both gravitation and refraction \label{s:4}}

For every gaseous refractive cloud characterized by a spherical symmetry, we obtain the static deviation angle $\Delta _{0}$ by 
means of equation (\ref{A7}) of Appendix \ref{s:A}. We schematically assume the possibilities to observe with ``present" or ``future" facilities, independently of the specific technological recipe. In the former case we have a a mirror of $D=1000$ cm and in the latter a mirror of $D=4000$ cm.  At $\lambda =5\times 10^{-5}$ cm, the relevant diffraction angle is $\theta _{d}\simeq 0.005$ and $0.004^{\prime \prime }$, respectively. The minimum detectable angle, being roughly $0.2$ $\theta _{d}$,
will therefore be $\vartheta _{Em\text{f}}=0.01$ and $0.0008^{\prime \prime }$, respectively. 

In the following subsections~\ref{s:4a} and \ref{s:4b} we apply those formulae to PSs and to MS stars. The aim is to predict the relevant microlensing. We also try to approximately compute the number of events that is clearly proportional to the number of cosmological sources that can be lensed (e.g. point-like sources as quasars), to the lenses, and to the cross-section of the latter. An evaluation of the probability of a microlensing event has however already been carried out in the past for cosmological sources as Gamma-Ray Bursts \citep[e.g.][]{1992ApJ...386L...5G,1993ApJ...402..382M,1998ApJ...495..597L} or quasars \citep{1973ApJ...185..397P,1995ApJ...455...37G}. For typical parameter this turns out to be close to unity depending also by the possibility that dark matter is at least partially composed by Massive Compact Halo Objects (MACHOs).

\subsection{Microlensings due to protostars in phase of star formation \label{s:4a}}

We here predict observable microlensings due to PSs, also calculating the number of them. Such number depends on the numbers of PSs
and quasars, and on the lensing cross-section. The latter one is expressed by:
\begin{equation}
\sigma _{C}=\pi \left( h_{M}^{2}-h_{m}^{2}\right) \text{ ,}  \label{24}
\end{equation}
where $h_{m}$ and $h_{M}$\ denote the minimum and maximum value, respectively, of the distances $h$ of minimum approach to the deviating PS.

The minimum value $h_{m}$ is limited by the opacity of the PS due for instance to the dust. A PS in the pre-main-sequence stage has typically eliminated the
majority of the dust surrounding it, although not completely as a main-sequence star \citep{2007ARA&A..45..565M}. Since we see the photosphere of a PS (having radius $\sim r_{C}$), the opacity cross-section $k_{\lambda }$ of a PS is roughly $2\times 10^{-8}$ that of the gas in our Galaxy, i.e.: 
\begin{equation}
k_{\lambda }\simeq 2\times 10^{-8}\times 2\times 10^{-22}[\text{cm}
^{2}]=4\times 10^{-30}[\text{cm}^{2}]\text{ }.  \label{eq:41}
\end{equation}

Denoting $h$\ a generic minimum approach, $dx$\ an adimensional differential such that $h$ $dx$\ represents a displacement, the intensity variation $dI$ after $h$ $dx$ can be expressed by: 
\begin{equation}
dI=-dx\text{ }h\text{ }k_{\lambda }NI\text{ ,}  \label{c}
\end{equation}
where, as usual, $N$\ denotes the gas number density. Integrating, we obtain: 
\begin{equation}
\ln \left( I_{0}\left/ I\right. \right) =h\text{ }k_{\lambda }\int_{-\infty
}^{+\infty }dx\text{ }N\left( x\right) \text{ ,}  \label{d}
\end{equation}
where $I_{0}$\ is the intensity value we would measure in absence of absorption.\ Since the deviation is very small, we approximate the
trajectory of the electromagnetic beam with a straight line along the $y$ axis. Setting: 
\begin{equation}
x=\tan \vartheta \text{ ,}  \label{e}
\end{equation}
it is $dx=d\vartheta $ $\cos ^{-2}\vartheta $, and the integral (\ref{d}), with the use of equation (\ref{9b}), turns out to be, after numerical
integrations and consequent interpolations: 
\begin{align}
& \ln \left( \frac{I_{0}}{I}\right) =hk_{\lambda }N\left( 0\right)
\int_{-\pi \left/ 2\right. }^{+\pi \left/ 2\right. }\frac{d\vartheta }{\cos
^{2}\vartheta }\left[ 1+\left( \frac{h}{r_{C}\cos \vartheta }\right) ^{2}
\right] ^{-1.8}  \notag \\
& =hk_{\lambda }N\left( 0\right) \left[ 1.77\left( \frac{r_{C}}{h}\right)
^{3.35}-1.17\left( \frac{r_{C}}{h}\right) ^{4}\right] \text{,}  \label{f}
\end{align}
valid for $0.8\leqslant h\left/ r_{C}\right. \leqslant 3.2$. The minimum $h_{m}\left/ r_{C}\right. $\ is obtained by imposing $\ln \left( I_{0}\left/
I\right. \right) =2$, which is a sensible choice between losing microlensing with a smaller $I\left/ I_{0}\right. $, and reducing the measurement time.
We obtain:
\begin{equation}
h_{m}=1.04r_{C}\text{ .}  \label{m}
\end{equation}

In order to find $h_{M}$, we first obtain the expression of the total static deviation due to both refraction and gravitation.\ According to our
procedure (see Sec. \ref{s:3}), we have to find the corresponding refraction indeces. In the case of a PS with core radius $r_{C}$, the refraction index
is given by equation (\ref{eq:10}), which can also be written as: 
\begin{equation}
n_{r}^{\text{PS}}\left( x\right) =1+5.28\times 10^{-7}\left(
1+x_{C}^{2}\left/ x^{2}\right. \right) ^{-1.8},  \label{eq:a13}
\end{equation}
having set 
\begin{equation}
x=h\left/ r\right. \text{ ,\ and \ }x_{C}=h\left/ r_{C}\right. \text{ ,}
\label{25}
\end{equation}
where $h$ denotes the distance of minimum approach.

\begin{figure}[tbp]
\begin{center}
\includegraphics[width=\columnwidth]{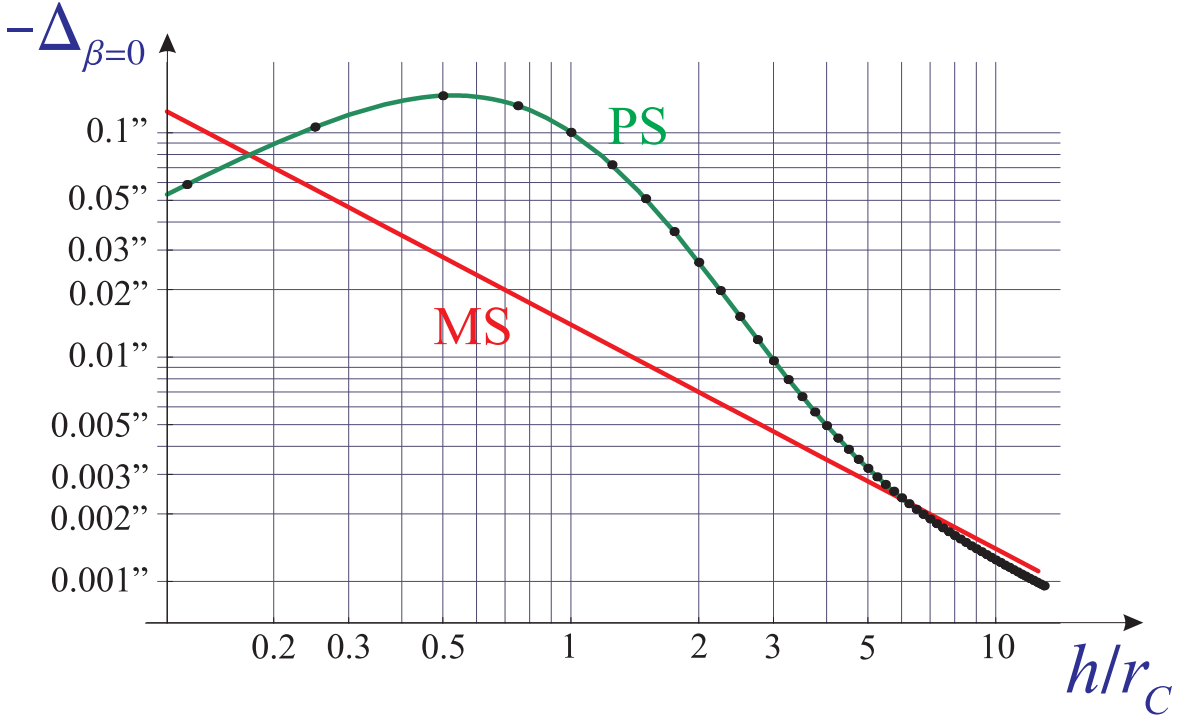}
\end{center}
\caption{The black circlets represent the deviation $\Delta _{\protect\beta =0}^{(PS)}$ due to a PS with mass $M_{\odot }$ in a frame at rest with the
PS \textit{vs} the ratio $h/r_{C}$, where $h$ is the distance of minimum approach, and $r_{C}$ the core of the King density profile. The
interpolating function is given by equation (\protect\ref{eqa14}) and represented in green. The red line shows the deviation $\Delta _{\protect
\beta =0}^{(MS)}$ due to the gravitational field of an MSS with mass $M_{\odot }$ and given by equation (\protect\ref{a}). The contribution due to
refraction of a MSS is nonnegligible only just outside the photosphere, as shown in the figure.}
\label{f1}
\end{figure}

As already said in the preceding sections, the gravitational deviation is
always present, and the relevant refractive index is given by equation (\ref
{p14}), where $\chi \left( r\right) $ is given by equations (\ref{p15}) and (\ref{p16}). In the latter one we write $\rho \left( r\right) =m_{p}N\left(
r\right) $, where $m_{p}$ denotes the proton mass, and $N\left( r\right) $ is given by equation (\ref{9b}). By means of numerical integrations followed
by interpolation, we obtain 
\begin{equation}
\mathbf{E}_{g}\left( \mathbf{r}\right) =-\mathbf{\hat{r}}\left( \frac{
2.37r_{C}}{r}+0.42+\frac{1.88r}{r_{C}}+\frac{0.61r^{2}}{r_{C}^{2}}\right)
^{-1}\left[ \frac{\text{cm}}{\text{s}^{2}}\right] \text{ .}  \label{e1}
\end{equation}
We derive from equations (\ref{p14}), (\ref{p15}), and (\ref{e1}): 
\begin{equation}
\chi \left( r\right) =6.73\times 10^{12}\frac{1+r\left/ r_{C}\right. }{
1+r\left/ r_{C}\right. +0.45r^{2}\left/ r_{C}^{2}\right. }\left[ \frac{\text{
cm}^{2}}{\text{s}^{2}}\right] \text{,}  \label{e2}
\end{equation}
and, with $x$ and $x_{C}$ as in equation (\ref{25}): 
\begin{align}
n_{G}^{\text{PS}}\left( x\right) & =\left[ 1+2.99\times 10^{-8}\frac{
1+x_{C}\left/ x\right. }{1+x_{C}\left/ x\right. +0.45x_{C}^{2}\left/
x^{2}\right. }\right] ^{1\left/ 2\right. }  \notag \\
& \simeq 1+1.49\times 10^{-8}\frac{1+x_{C}\left/ x\right. }{1+x_{C}\left/
x\right. +0.45x_{C}^{2}\left/ x^{2}\right. }\text{.}  \label{e3}
\end{align}
In presence of gravitation, the light speed $c_{\infty }$ reduces in vacuum
to $c=c_{\infty }\left/ n_{G}\right. $, and in a medium to: 
\begin{equation}
v_{L}=\frac{c}{n_{r}}=\frac{c_{\infty }}{n_{r}n_{G}}=\frac{c_{\infty }}{n}
\text{,}  \label{e4}
\end{equation}
where: 
\begin{equation}
n=n_{r}n_{G}\text{.}  \label{e5}
\end{equation}
Substituting equation (\ref{e5}) into (\ref{A7}) of Appendix \ref{s:A}, we obtain the values for the deviations $\Delta _{0}$\ shown in Fig.\textbf{\
\thinspace \ref{f1}} for different $x_{C}=h/r_{C}$ ratios, which are well interpolated by: 
\begin{equation}
\Delta _{0}^{\text{PS}}=\frac{0.333x_{C}-0.0655x_{C}^{2}+0.0155x_{C}^{3}}{0.684-0.163x_{C}+1.69x_{C}^{2}-0.404x_{C}^{3}+x_{C}^{4}}  \label{eqa14}
\end{equation}
measured in arcseconds.

Actually, we do not observe $\Delta _{0}^{(PS)}$, but the angle $\vartheta_{E}$ between the non deviated ray and the deviated ray near the Earth,
related to \ $\Delta _{0}^{(PS)}$ through Eq. (\ref{53}). Refrerring to ``present" and ``futures" facilities the minimum detectable $\Delta _{zm}^{(PS)}$\ is therefore given by
\begin{equation}
\Delta _{zm\text{p}}^{(PS)}=\frac{1+z+z^{2}/2}{\sqrt{1+z+z^{2}}}\text{ }
0.01^{\prime \prime }\text{ ,}  \label{54}
\end{equation}
and by
\begin{equation}
\Delta _{zm\text{f}}^{(PS)}=\frac{1+z+z^{2}/2}{\sqrt{1+z+z^{2}}}\text{ }0.0008^{\prime \prime }\text{ .}  \label{55}
\end{equation}
For $z=0$, the $z$ term appearing in Eqs. (\ref{54}) and (\ref{55}) becomes unity. The maximum static value $h_{M0}$ of \ the distance of minimum
approach to the PS can be derived from equation (\ref{eqa14}), or more easily from Fig. \textbf{\thinspace \ref{f1}, }in correspondence of the
minimum static deviations. With present and future facilities, we obtain
\begin{equation}
h_{M0\text{p}}(0.01^{\prime \prime })=2.96\text{ }r_{C}\text{ ; }h_{M0\text{f
}}(0.0008^{\prime \prime })=16.2\text{ }r_{C}\text{ .\ }  \label{56}
\end{equation}
The corresponding cross-sections can be derived from equations (\ref{24}),
(\ref{m}), and (\ref{56}):
\begin{equation}
\sigma _{C\text{p}}=7.68\text{ }\pi \text{ }r_{C}^{2}\text{ , and }\sigma _{C
\text{f}}=261.4\text{ }\pi \text{ }r_{C}^{2}\text{ .}  \label{57}
\end{equation}
We see that the ratio between the future and present cross-section is $34$, and such ratio is roughly preserved for $z>0$, the number of microlensings
due to PSs and detectable with future facilities is $34$\ times the one with present facilities. 

\subsubsection{Number of microlensing events}

In order to calculate the number $\mathcal{N}_{PS}(\beta _{k})$ of PSs we start from the rate of star formation of \citet{2008MNRAS.388.1487L}, given by:
\begin{equation}
SFR(z)=\frac{0.0157+0.118z}{1+(z/3.23)^{4.66}}\text{ [M}_{\odot }\text{yr}
^{-1}\text{Mpc}^{-3}\text{] .}  \label{40}
\end{equation}

The number $\mathcal{N}_{PS}(\beta _{k})$ can be derived
from equation (\ref{40}):
\begin{equation}
\mathcal{N}_{k}^{\text{PS}}=\frac{4}{3}\pi R_{H\text{[Mpc]}
}^{3}t_{PS}\int_{\beta _{k-1}}^{\beta _{k}}d\beta \text{ }\beta ^{2}\frac{0.0157+0.118z(\beta )}{1+[z(\beta )/3.23]^{4.66}}\text{ ,}  \label{41a}
\end{equation}
where $R_{H\text{[Mpc]}}=4.23\times 10^{3}$\ denotes the Hubble radius measured in Mpc, $t_{PS}$\ $\simeq 4\times 10^{7}$ [yr] the duration of the
PS stage \citep{2007ARA&A..45..565M}, and, as derived from equation (\ref{50}), 
\begin{equation}
z(\beta )=\sqrt{\frac{1+\beta }{1-\beta }}-1\text{ .}  \label{42}
\end{equation}

We express the cross-sections through:
\begin{equation}
y_{k}=\sigma _{C\text{f}}(\bar{\beta}_{k})/\pi r_{C}^{2}\text{ ,}  \label{45}
\end{equation}
where $\bar{\beta}_{k}$\ is the average $\beta $ value between $k-1$\ and $k$, and $\sigma _{C\text{f}}$ can be derived from equations (\ref{24}), (\ref{m}), (\ref{25}), and (\ref{eqa14}).

\bigskip The number of quasars that can produce microlensing can be potentially very large. \citet{2011ApJS..195...10X} estimated in deep {\it Chandra} images a density of about $10^4$ Active Galactic Nuclei (AGN) per square degree. How many of them are suitable for a monitoring depends on optical brightness and redshift distribution and analytical estimates \citep[e.g.][and references therein]{1994ApJ...424..550D} shows that the optical depths for lensing is easily close to one if the survey includes sources at sufficiently high redshift ($Z \geqslant 3$). For simplicity, and because the computation has points of interest, we follow here a simpler approach and the number of quasars that can produce a microlensing is given by:
\begin{equation}
\mathcal{N}_{k+1}^{\text{Q}}=\int_{\beta _{k+1}}^{0.98}\mathcal{N}_{Q}(\beta
)\text{ }d\beta \text{ ,}  \label{46}
\end{equation}
where the number of quasars present at a given age, here characterized by $\beta$, is obtained by the known value of $1$\ quasar in $160$\ galaxies
for $\beta =0.2$, and the sensible hypothesis that, for $\beta \leq 0.98$, one of three galaxy cores was a quasar, and, for $\beta >0.98$ there is such
a rapid decrease of quasar density, that we\ can consider it equal to zero.
Under that assumption we obtain:
\begin{equation}
\mathcal{N}_{Q}(\beta )=10^{11}\exp \left( \frac{\beta -0.98}{5.08}\right) 
\text{ ,}  \label{47}
\end{equation}
so that we derive from equations (\ref{46}) and (\ref{47}):
\begin{equation}
\mathcal{N}_{k+1}^{\text{Q}}=0.5\times 10^{12}\left[ 1-\exp \left( \frac{\beta _{k+1}-0.98}{5.08}\right) \right] \text{ .}  \label{48}
\end{equation}

Finally, the number of microlensing due to PS in each shell centered on $k$
is given by:
\begin{equation}
\mathcal{N}_{\text{mic,}k}^{\text{PS}}=\mathcal{N}_{k}^{\text{PS}}\mathcal{N}
_{k+1}^{\text{Q}}\frac{y_{k}r_{C}^{2}}{4\beta _{k}^{2}R_{H}^{2}}\text{ .}
\label{49a}
\end{equation}

\begin{figure*}[tbp]
\begin{center}
\begin{tabular}{cc}
\includegraphics[width=\columnwidth]{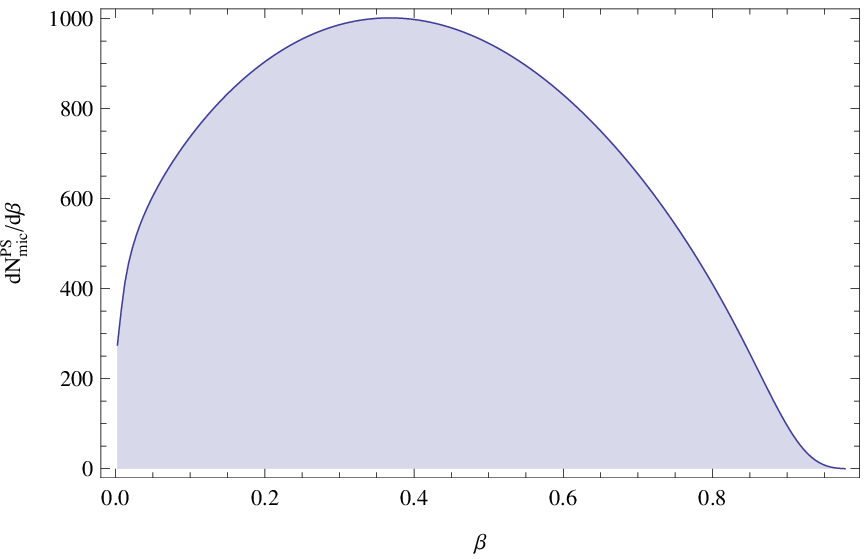} & \includegraphics[width=\columnwidth]{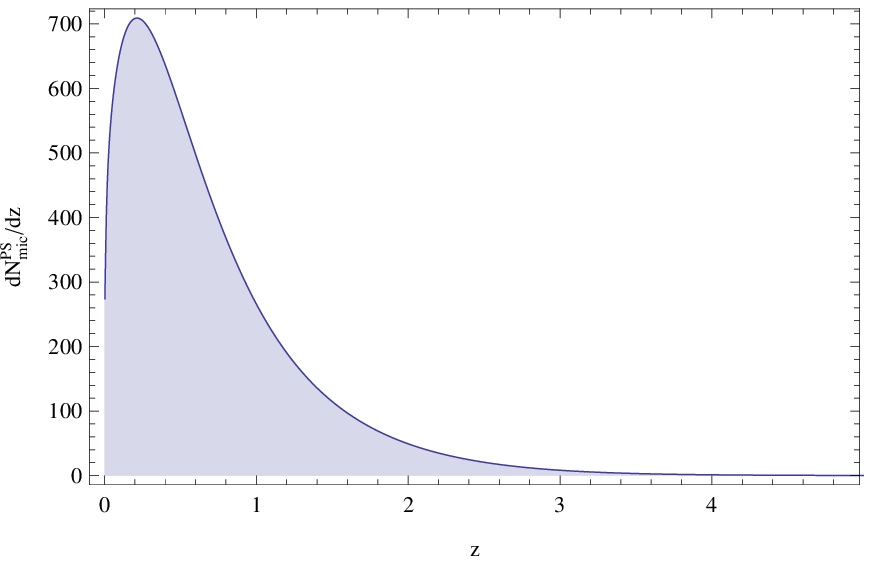}
\end{tabular}
\end{center}
\caption{{\it (left)} The number of of microlensing due to PSs as a function of $\beta$ and {\it (right)} the same number as a function of the redshift $z$.}
\label{inbetazeta}
\end{figure*}

We report $d\mathcal{N}_{\text{mic}}^{\text{PS}}/d\beta $ in Fig(s). \ref{inbetazeta}, whence, by integrating over $\beta $\ from $\beta _{1}$\ to $\beta
_{2}$, we obtain the number of observable\ microlensings in the interval $\beta _{2}-\beta _{1}$. Similarly, we report $d\mathcal{N}_{\text{mic}}^{\text{PS}}/dz$ in the same figure, whence we can obtain the number of observable\ microlensings in the interval $z_{2}-z_{1}$.

The integration over both full intervals of Fig(s). \ref{inbetazeta} obviously gives the same result that is about $700$ observable
microlensings. That results, as expected, implies that the phenomenum should be fairly common and potentially observable through dedicated surveys. 

As a byproduct, we may confirm that the lens effect is little affected by the refractions of the PS (and, to a greater reason, by the
refraction of other astrophysical objects). In fact, the microlensing can give at maximum a deviation angle of $\sim0.02^{\prime\prime}$, while the
gravitational lens effects is between $0.8^{\prime\prime}$ and $5^{\prime \prime}$.

\subsection{Microlensings due to main-sequence stars \label{s:4b}}

For MSSs we only consider the gravitational deviation because the one due to refraction in the considered range $0.001^{\prime \prime }<\Delta
<0.1^{\prime \prime }$ has a negligible probability to be observed.\ In fact, for a star having the Sun's mass, we derive from equation (\ref{eq:47}) and the data taken from the Sun 
\begin{equation}
N\simeq 1.3\times 10^{17}\exp \left( \frac{r_{P}-r}{1.8\times 10^{5}\left[ 
\text{m}\right] }\right) \ \left[ \text{c}\mathrm{m}^{-3}\right] \ ,
\label{eq:x}
\end{equation}
for $r>r_{P}$, and with $r_{P}=7.0\times 10^{10}$ cm. In the case of a MSS, the gas, whose density is sufficient to produce an appreciable deviation, is ionized, and $n_{r}^{(MS)}$ is derivable from equations (\ref{eq:1})-(\ref{eq:3}), leading to: 
\begin{align}
n_{r}^{(MSS)}\left( r\right) & =1-\frac{e^{2}N\left( r\right) }{2\omega
_{y}^{2}m_{e}\varepsilon _{0}}  \notag \\
& =1-2.86\times 10^{-5}\exp \left( \frac{r_{P}-r}{1.8\times 10^{5}\left[ 
\text{m}\right] }\right) \ \ ,  \label{eq:y}
\end{align}
for $r>r_{P}$. Substituting equation (\ref{eq:y}) into equation (\ref{A7}) of Appendix \ref{s:A}, we find the deviations (only due to refraction)\ for
different $h/r_{p}$ values (where $h$ denotes the distance of minimum approach). Interpolating the results, we obtain: 
\begin{equation}
\Delta _{\beta =0}^{(MSS)}=2499^{\prime \prime }\exp \left[ -3896\left( \frac{
h}{r_{p}}-1\right) \right] \text{ ,}  \label{a}
\end{equation}
for $1.0012<h/r_{p}<1.004$.

In the range $-0.1^{\prime \prime }\leqslant \Delta_{refractive}^{(MSS)}\leqslant -0.001^{\prime \prime }$, we have $h_{m}\leqslant h\leqslant h_{M}$, with: 
\begin{equation}
h_{m}=1.00260r_{p}\text{ \ and \ }h_{M}=1.00378r_{p}\text{ .}  \label{b}
\end{equation}

The useful cross-section to have a deviation of light between $0.001^{\prime
\prime }$ and $0.1^{\prime \prime }$ is therefore: 
\begin{align}
\sigma _{star}& =\pi \left[ \left( r_{p}+h_{M}\right) ^{2}-\left(
r_{p}+h_{m}\right) ^{2}\right] \simeq 2\pi r_{p}(h_{M}-h_{m})  \notag \\
& =2.4\times 10^{18}\text{ [cm}^{2}\text{] \ ,}\   \label{cd}
\end{align}
where we have taken $r_{p}=R_{\odot }=7.8\times 10^{8}$m. The ratio of this cross-section and the one relevant to gravitation is given, with the use of
equations (\ref{9}), (\ref{57}), and (\ref{cd}), by $3.8\times 10^{-11}$.
The microlensing due to MSSs is therefore due to gravitation only, which is comparable with the one due to PS.

\section{Conclusions \label{s:5}}

The relative abundance of ionized hydrogen is high in the universe, so that we have examined the possibility that it may cause measurable deviations of rays of light because of refraction. Dense ionized hydrogen is confined in the H II regions whose projected area is only $\sim2\%$ of the whole galaxy
projected area, and are close to the central part of the plane of symmetry of the galaxy \citep{1984ApJ...276...38J,1999ApJ...511...65J}. Ionized hydrogen is of course diffused in the universe but with very low density. The refraction index of the H II
region plasma is so close to unity that no appreaciable deviation can ever be detected. The ionized gas in the galaxy clusters has a still smaller
density, hence a still smaller plasma frequency. Consequently, the corresponding deviations are still smaller than those due to the H II regions. The only regions of ionized hydrogen where a strong refraction occurs are the star shells beyond the star photospheres. However, the probability that the light coming from a distant quasar traverses the effective area of a star atmosphere is negligible.

Passing to clouds of neutral hydrogen, galaxies, giant clouds, and dense clouds give negligible deviations. The only clouds that give observable deviations due to refraction, are the protostars in the pre-main sequence stage.

The application to the lens effect, in order to see possible corrections to the main gravitational deviation, has been performed in Sect.\,\ref{s:3}, and applied to spherical clouds of neutral hydrogen. The result for the deviation $\Delta$ of a beam of light is given by equations (\ref{54}) and (\ref{55}), and depends on $\varepsilon =n_{ry}-1$ (where $n_{ry}$\ is the refractive index for yellow light), $\beta=v/c$ (where $v$ is the cosmological recession speed), the radius $R$ of the core of the object, and $h$ the impact parameter of the light beam with respect to the center of the object. The obtained value for the
deviation is $\Delta=|\delta-\varepsilon|\simeq5\times10^{-8}$ rad $\simeq10^{-2\ ^{\prime \prime}}$, which is $\sim10\%$ the average observable deviation due to the gravitational lens effect. Therefore 
the additional deviation due to refraction by part of the intergalactic gas is essentialy negligible.

In Sec.\,\ref{s:4} we have evaluated the probability that one of the two light beams constituting the lens effect is appreciably deviated by the atmosphere of a star belonging to the considered receding galaxy. The probability of such an event turns out to be negligible, as can be seen from equations (\ref{cd}). Concluding, the additional deviation due to relativistic refraction is negligible, and possible deviations of some arc seconds due to stellar atmospheres is a very rare event.

The second, and main, contribution of our work is the possibility to observe microlensing due to PSs. In Sec. \ref{s:4a}, we have found that the static
deviations are in the range $0.001^{\prime \prime }<\Delta _{\beta=0}<0.012^{\prime \prime }$, measurable with modern facilities, and that the
total number $\mathcal{N}_{PS}^{\text{tot}}$ of observable microlensings due to PSs is high enough to make a deoved survey feasible. The most important information could be obtained by microlensings due to PSs in high red-shift receding galaxies. Indeed, the visibility of a quasar, whose light traverses a PS cloud reaching a minimum
distance $h\simeq 0.8r_{C}$, is possible for near PSs if using $\lambda \simeq 2.2\mu m$. However, if the quasar light traverses a PS gas cloud receding with high redshift, observation at $\lambda \simeq 2.2\mu m$ implies that the wavelength $\lambda $ was in the visible if observed in a frame fixed to the considered PS, potentially even drawing information about the dust percentage vs $z$.

\appendix

\section{Calculation of the deviation due to the refraction of a wave beam traversing a gas cloud with spherical symmetry}

\label{s:A}

In a frame at rest with a cloud of gas having refraction index $n(\mathbf{r}) $, function of $r=\left\vert \mathbf{r}\right\vert $, the deviation of a
wave beam can be obtained by an integral, as shown in Appendix B of \citet{2013PhRvE..87d3202C}. The resulting expression was: 
\begin{equation}
\varphi _{\mathbf{B}}-\varphi _{\mathbf{A}}=\int_{\mathbf{A}}^{\mathbf{B}}
\frac{dr}{r}\left( \frac{n^{2}r^{2}}{G^{2}}-1\right) ^{-1/2}\ .  \label{A1}
\end{equation}
where $G$\ denotes the absolute value of the Bouguer vector: 
\begin{equation}
G=\left\vert \mathbf{G}\right\vert =\left\vert n\left( r\right) \mathbf{r}
\times d\mathbf{r}\left/ ds\right. \right\vert \text{ ,}  \label{A2}
\end{equation}
$s$ denoting the curviliner abscissa. If we want the total variation of the $\varphi $ angle when the wave ray starts from, and then goes to infinity, it
is convenient to calculate the deviation of light for a ray coming from infinity and ending at the distance of minimum approach $h$, where the
Bouguer vector reduces to: 
\begin{equation}
G=n\left( h\right) hdr/ds=n\left( h\right) h\text{ .}  \label{A3}
\end{equation}
The wanted deviation is then obtained by doubling the result due to the considered half path\textsl{\ }\citep{2013PhRvE..87d3202C}: 
\begin{equation}
\delta \varphi _{tot}=2\int_{1}^{+\infty }\frac{dr}{r}\left[
\left( \frac{n\left( r\right) r}{n\left( h\right) h}\right) ^{2}-1\right]
^{-1/2}\text{ .}  \label{A4}
\end{equation}
Setting $x=r\left/ h\right. $ in the last integral, Eq. (\ref{A4}) becomes: 
\begin{equation}
\delta \varphi _{tot}=2\int_{1}^{+\infty }\frac{dx}{x}\left[ \left( x\frac{
n\left( x\right) }{n\left( 1\right) }\right) ^{2}-1\right] ^{-1/2}\text{ .}
\label{A5}
\end{equation}
When $n\left( x\right) =1$ (negligible refraction), it is: 
\begin{equation}
\delta \varphi _{tot}\left( n=1\right) =2\int_{1}^{+\infty }\frac{dx}{
x\left( x^{2}-1\right) ^{1/2}}=\pi \ .  \label{A6}
\end{equation}

The total deviation $\Delta_{\beta=0}$ of the wave ray coming from the quasar, passing at a minimum distance $r_{0}$ from the center of a gas
cloud, and arriving at the Earth is: 
\begin{align}
\Delta_{\beta=0} & =\delta\varphi_{tot}\left( n=1\right) -\delta
\varphi_{tot}=  \notag \\
& =\pi-2\int _{1}^{+\infty}\frac{dx}{x}\left[ \left( x\frac{n\left(
x\right) }{n\left( 1\right) }\right) ^{2}-1\right] ^{-1/2}\ ,  \label{A7}
\end{align}
with the symbol \textquotedblleft$\beta=0$" indicating the non-relativistic calculus, i.e., the choice of a reference system at rest with the gas cloud.

\acknowledgments
This work has been supported by ASI grant I/004/11/2.

\bibliographystyle{aa}
\bibliography{Microlen}

\end{document}